\documentclass[
    ,final            
  ]
  {aipproc}

\layoutstyle{8s}

\begin{document}

\title{Indications for the onset of deconfinement \\ in nucleus nucleus collisions}

\classification{25.75.Nq}
\keywords      {Quark Gluon Plasma, Heavy Ion Collisions, Phase Transition, NA49}


\author{
{\small
D.~Flierl$^{9}$,C.~Alt$^{9}$, T.~Anticic$^{21}$, B.~Baatar$^{8}$,D.~Barna$^{4}$,
J.~Bartke$^{6}$, 
L.~Betev$^{9,10}$, H.~Bia{\l}\-kowska$^{19}$, A.~Billmeier$^{9}$,
C.~Blume$^{9}$
% ...
}\\
\vspace{0.2cm}
The NA49 collaboration}{
address={
\small{
% ...
$^{4}$KFKI Research Institute for Particle and Nuclear Physics, Budapest, Hungary.\\
% ...
$^{6}$Institute of Nuclear Physics, Cracow, Poland.\\
% ...
$^{8}$Joint Institute for Nuclear Research, Dubna, Russia.\\
$^{9}$Fachbereich Physik der Universit\"{a}t, Frankfurt, Germany.\\
$^{10}$CERN, Geneva, Switzerland.\\
% ...
$^{19}$Institute for Nuclear Studies, Warsaw, Poland.\\
% ...
$^{21}$Rudjer Boskovic Institute, Zagreb, Croatia.
}
}
}

\begin{abstract}
 The hadronic final state of central Pb+Pb collisions at 20, 30, 40, 80, and 158 AGeV has been measured by the CERN NA49 collaboration. 
The mean transverse mass of pions and kaons at midrapidity stays nearly constant in this energy range, whereas at 
lower energies, at the AGS, a steep increase with beam energy was measured.
Compared to p+p collisions as well as to model calculations, anomalies in the energy dependence of pion and kaon production at lower SPS energies are observed.
These findings can be explained, assuming that the energy density reached in central A+A collisions at lower SPS energies is sufficient
to force the hot and dense nuclear matter into a deconfined phase.
\end{abstract}

\maketitle

\section{Introduction}
At top CERN SPS and at RHIC energies the initial energy density reached in heavy ion collisions is apparently large enough to create a deconfined phase 
at the early stage of the collision: the Quark Gluon Plasma \cite{Heinz:2000bk}\cite{Gyulassy:2004zy}. 
On the other hand, at lower energies, at the AGS, the maximum energy density is 
probably not sufficient to produce such a state of matter. 
Therefore, an energy scan progam in the intermediate energy range, covered by the CERN SPS, was initiated. 
The goal was to find anomalies in the energy dependence which could be related to the change of the number of degrees of freedom 
connected to a phase transition of the hot and dense matter created in heavy ion collisions.

In the course of the SPS energy scan program, the NA49 collaboration measured the hadronic final state of central Pb+Pb collisions 
at 20, 30, 40, 80, and 158 AGeV beam energy. A detailed description of the NA49 setup is given in \cite{Afanasev:1999iu}.
In the following, the energy dependence of selected hadronic observables will be discussed,
with the main focus on where we start to see indications of a deconfined phase of matter.

\section{Energy Dependence of Transverse Mass Spectra}

A model independent way to compare transverse momentum spectra is to consider the mean transverse mass 
$\langle{m_t}\rangle - m_0$.
In Figure 1 the energy dependence of this quantity is shown for pions, kaons and protons.

If energy density and hence pressure increase with beam energy, a stronger transverse expansion is
expected at higher energies.  Assuming that the strength of the transverse expansion is reflected in the mean transverse mass, 
the observable $\langle{m_t}\rangle - m_0$ will rise with $\sqrt{s_{NN}}$.

\begin{figure}[h]
\includegraphics[height=5cm]{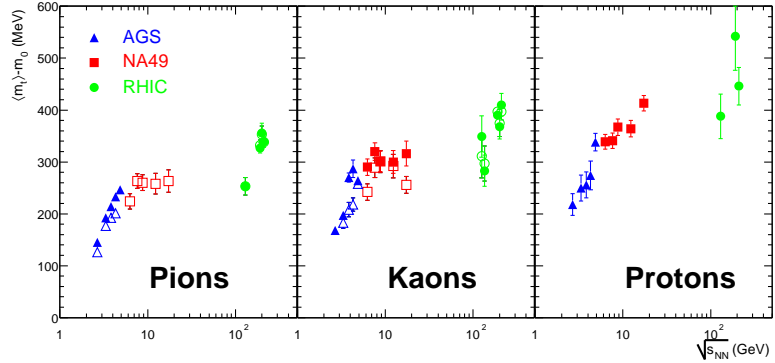}
\caption{Energy dependence of the mean transverse mass of pions, kaons and protons. Open symbols indicate negatively charged particles.}
\end{figure}

At AGS energies we indeed observe a strong increase of the mean transverse mass of pions, kaons and protons with energy. However, at lower SPS energies this behaviour
changes: the mean transverse mass increases only weakly with beam energy.

This charateristic change in the energy dependence has been interpreted as a signature for a phase transition \cite{Gazdzicki:1998vd}. Because
the equation of state - the relation between energy density and pressure - changes at the phase boundary and with it the strength of the transverse expansion.

This observation supports the hypothesis, that the initial energy density reached in central A+A collisions at lower SPS energies is already large enough, 
to force a phase transition of the strongly interacting nuclear matter. 

\section{Energy Dependence of Particle Yields}

\begin{figure}[h]
\includegraphics[height=4.9cm]{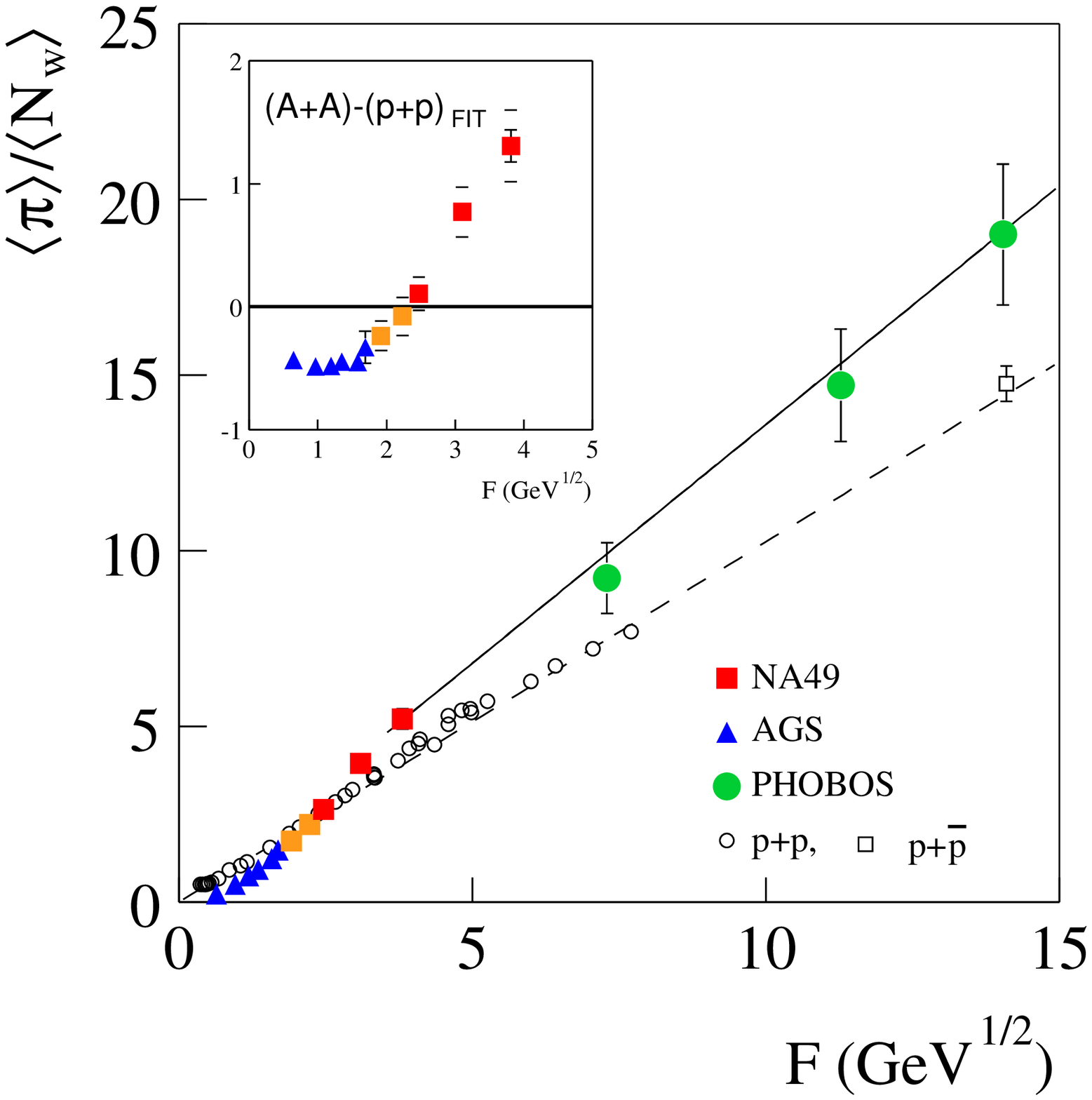}
\includegraphics[height=4.9cm]{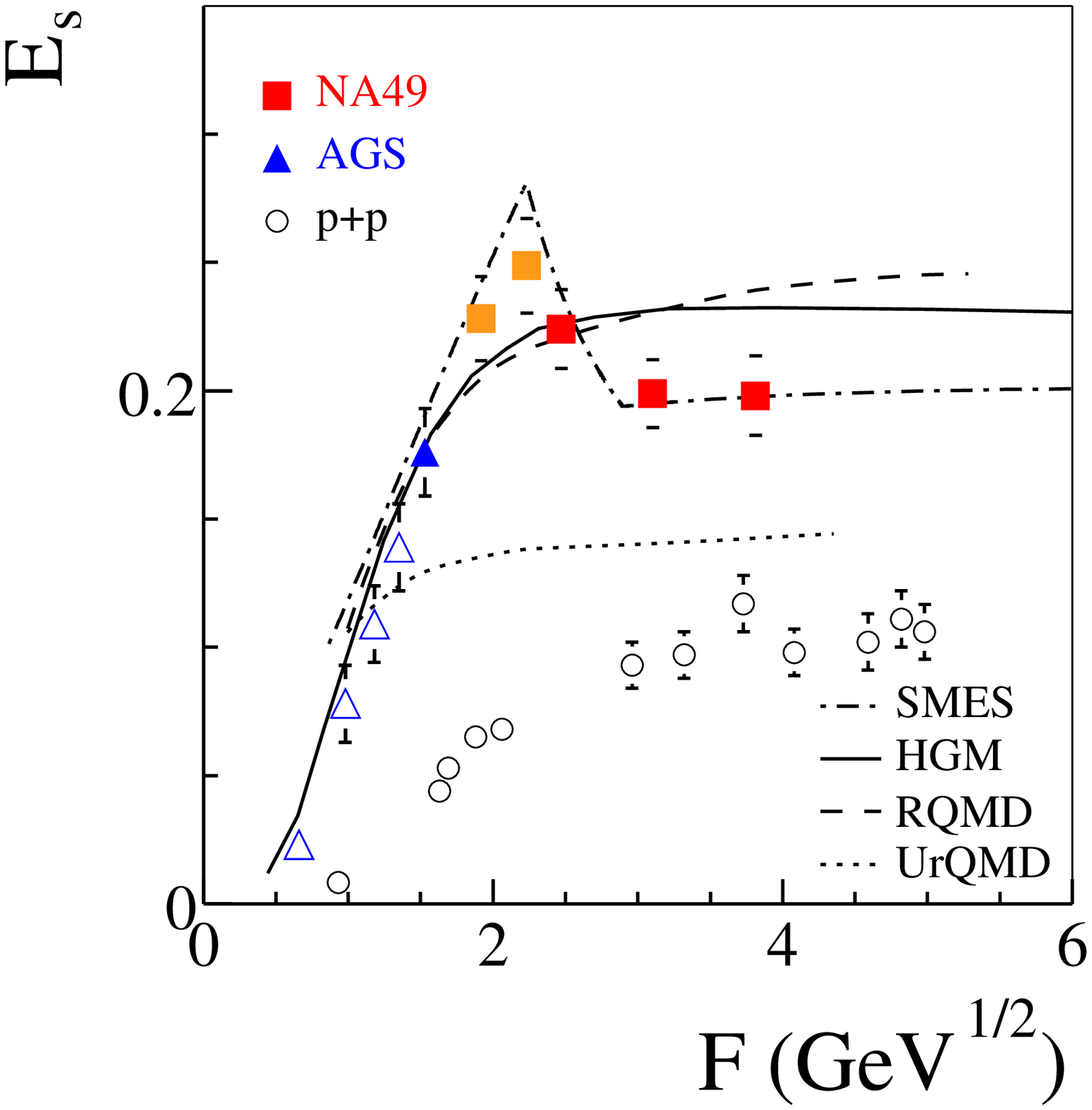}
\caption{Energy dependence of the mean number of pions per wounded nucleon in A+A and p+p collisions. Lines are only drawn to guide the eye (left panel).\newline
Energy dependence of the strangeness to pion ratio and model calculations (right panel).}

\label{fig:rapidity}
\end{figure}

The total pion multiplicity per wounded nucleon as a function of Fermi's measure $F \approx \sqrt{\sqrt{s_{NN}}}$ is shown in Figure 2 left. 
The relative pion production in central Pb+Pb collisions increases with collision energy, but starting from lower SPS energies the rate of increase grows. 
This feature is not visible in p+p interactions. 
The steepening of the energy dependence can be explained by an increase of the effective number of degrees of freedom due to the onset of deconfinement at
about 30 AGeV beam energy \cite{Gazdzicki:1998vd}.
         
Figure 3 shows the energy dependence of relative strangeness production. The ratios of the total multiplicities ${\langle \rm K^+ \rangle} / {\langle\pi^+\rangle}$, 
${\langle \rm K^- \rangle} / {\langle\pi^-\rangle}$, 
${\langle \Lambda \rangle} / {\langle\pi\rangle}$  and ${\langle \overline \Lambda \rangle} / {\langle\pi\rangle}$ in central Pb+Pb collisions are compared to model 
calculations and when available to p+p interactions. 
While  ${\langle \rm K^- \rangle} / {\langle\pi^-\rangle}$ and ${\langle \overline \Lambda \rangle} / {\langle\pi\rangle}$ ratios rise continuously with energy,
 a distinct maximum is visible in the energy dependence of the ratios ${\langle \rm K^+ \rangle} / {\langle\pi^+\rangle}$ and ${\langle \Lambda \rangle} / {\langle\pi\rangle}$.
Hadron gas \cite{Cleymans:1999st} (HGM) and microscopic models \cite{Weber:2002pk} (RQMD, UrQMD) roughly describe the trend of the energy dependence, but especially the pronounced peak in 
the ${\langle \rm K^+ \rangle} / {\langle\pi^+\rangle}$ ratio is not reproduced. 
 As demonstrated in Figure 3, this feature is also absent in p+p interactions.    

\begin{figure}[h]
\includegraphics[height=4.0cm]{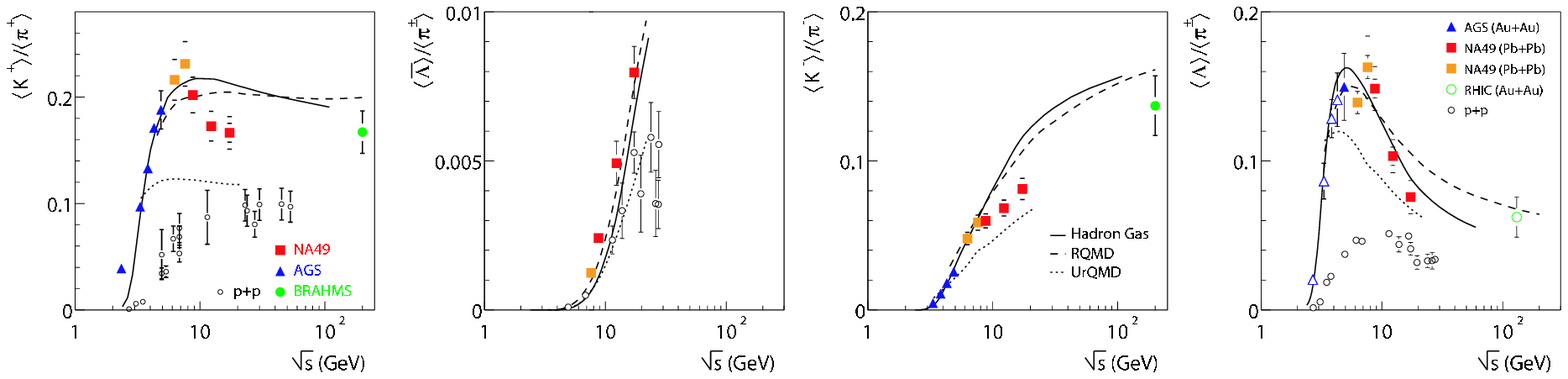}
\caption{Energy dependence of particle ratios  ${\langle \rm K^+ \rangle} / {\langle\pi^+\rangle}$, ${\langle \overline \Lambda \rangle} / {\langle\pi\rangle}$,
${\langle \rm K^- \rangle} / {\langle\pi^-\rangle}$, and 
${\langle \Lambda \rangle} / {\langle\pi\rangle}$ in central A+A collisions and p+p interactions.
Lines indicate model calculations.}
\label{fig:strangeness}
\end{figure}

The total strangeness production to pion rate can be approximated by $E_S = (2(\langle \rm K^- \rangle + \langle \rm K^- \rangle) + \langle \Lambda \rangle) / \langle \pi \rangle$.
As seen from Figure 2 (right), we find that the hadron gas and microscopic models describe the trend of the data, but they do not show the 
distinct maximum at lower SPS energies. Only a model \cite{Gazdzicki:1998vd} (SMES) including a phase transition around 30 AGeV  exhibits a peak as seen in the data.

\section{Summary and Outlook}
The energy dependence of several observables shows anomalies at lower SPS energies. These signatures indicate, that the energy density reached in A+A collisions
at beam energies of about 30 AGeV are already sufficient to produce a deconfined phase.

The next question to address is, how the system behaves when the volume is varied. At which intermediate system size appear indications for a deconfined phase? 
Future experiments at the SPS could deliver the answer to this question if the proposed light ion program will be approved \cite{Letter:2003aa}.

\begin{theacknowledgments}
This work was supported by the US Department of Energy
Grant DE-FG03-97ER41020/A000,
the Bundesministerium fur Bildung und Forschung, Germany, 
the Polish State Committee for Scientific Research (2 P03B 130 23, SPB/CERN/P-03/Dz 446/2002-2004, 2 P03B 04123), 
the Hungarian Scientific Research Foundation (T032648, T032293, T043514),
the Hungarian National Science Foundation, OTKA, (F034707),
the Polish-German Foundation, and the Korea Research Foundation Grant (KRF-2003-070-C00015).
\end{theacknowledgments}

\end{document}